\documentstyle[aas2pp4]{article}
\lefthead{Reale et al.}
\righthead{Temperature and emission measure of filamented TRACE loops}
\begin{document}

\title{TRACE-derived temperature and emission measure profiles along 
long-lived coronal loops: \\ the role of filamentation.}

\author{F. Reale\footnote{E-mail: {\it reale@oapa.astropa.unipa.it}}, 
G. Peres\footnote{E-mail: {\it peres@oapa.astropa.unipa.it}}}

\affil{Dip. di Scienze Fisiche \& Astronomiche -- Sez. di
Astronomia -- \\ Univ. di Palermo, Piazza del Parlamento 1, I--90134
Palermo, Italy}

\begin{abstract}
In a recent letter (ApJ 517, L155) Lenz et al. have shown the evidence
of uniform temperature along steady long coronal loops observed by
TRACE in two different passbands (171 \AA~ and 195 \AA~ filters).  We
propose that such an evidence can be explained by the sub-arcsecond
structuring of the loops across the magnetic field lines.  In this
perspective, we present a model of a bundle of six thin parallel
hydrostatic filaments with temperature stratification dictated by
detailed energy balance and with temperatures at their apex ranging
between 0.8 and 5 MK. If analyzed as a single loop, the bundle would
appear isothermal along most of its length.
\end{abstract}


\section{Introduction}
\label{sec:intro}

The solar X-ray emitting corona entirely consists of independent
loop-like bright structures, in which the plasma is confined by the
magnetic field (e.g. Vaiana et al. 1973). The coronal loops cover a
wide range of sizes and brightness, and make the solar corona highly
structured and contrasted in X-rays. Observations at high angular
resolution (1") made with the {\it Transition Region and Coronal
Explorer} ({\it TRACE}, e.g. Handy et al. 1999) show, once again, the
high level of structuring of the solar corona, and, in particular, that
coronal loops typically consist of several individual filaments, down
to the telescope resolution limit. This raises even more questions on
the structure, evolution, interaction, turning on, coherence and
eventually the heating of the individual filaments (e.g. Proceedings of
the Workshop on the Physics of the Solar Corona and Transition Region,
24 - 27 August 1999 Monterey, California, Solar Physics, 1999, in
press). In particular, one basic question is how the physical
conditions of the several, energetically independent, loop filaments
are related to the physical parameters derived from the analysis of
TRACE data.

The temperature and density distribution along some loop structures observed
by TRACE has been recently investigated by Lenz et al. (1999, hereafter
L99). They have selected four relatively isolated loops extending above the
solar limb and steady, at least for time intervals of 1-2 hr. The half
length of three of the loops (loops a, b, and d) is $L \approx 10^{10}$
cm, and that of the fourth (loop c) is $L \approx 5 \times
10^9$ cm. L99 have analyzed the brightness distribution along
each loop by selecting four subimages of each loop (at 1/5, 1/3, 2/3 and
3/3 of the distance from the base to the loop top). Each subimage
contains a few hundred to a few thousand pixels and includes the whole
loop cross-section.

The temperature and the emission measure (EM) in the regions mentioned above
are derived from the 171/195 \AA~ filter ratio and from the 171 \AA~
count rate, respectively, by assuming that all the plasma along the line
of sight is at the same temperature and density. L99 find that the
filter ratio varies very little along the four loops, therefore
concluding that the temperatures profiles are constant along the loops,
around 1.3 MK. This finding is at variance from a static, steady-state,
non-isothermal loop model (e.g. Serio et al. 1981), uniformly heated
and with a temperature of 1.3 MK at the apex: although the temperature
profile would be rather flat in the corona according to this model, the
observed profile is significantly (and incompatibly) flatter.

By assuming a uniform line-of-sight depth of $10^{10}$ cm, L99 obtain
emission measures increasing from the apex to the base, and ranging
between $10^{27}$ and $10^{28}$ cm$^{-5}$. Such profiles are flatter,
and at higher density, than predicted by single loop hydrostatic models
at that temperature (Serio et al. 1981), and are better described by
isothermal loops at T=1.3 MK.

Although aware that the static non-isothermal models used do not take
into account possible additional effects, such as non-uniform
heating, the presence of flows and mixing, and wave interaction with
the background fluid, L99 conclude that "the lack of temperature variation
in the EUV loops considered invites speculation that there is a class
of such isothermal loops distinct from loops with a temperature maximum
at the apex."

This letter revisits this interpretation in light of a more realistic
scenario in which the loops detected by TRACE consist of bundles of
filaments independent of each other, possibly in different physical
conditions, and, in particular, at different temperatures.  Each
filament may then be described by a distinct non-isothermal loop model.
We will show that TRACE diagnostic may actually yield an apparently
almost isothermal profile along most of a loop consisting of such a
bundle.

We conclude, therefore, that the evidence of isothermal loops shown by
TRACE may just be a further signature of the filamentary structure of
coronal loops and of the multi-temperature nature of each bundle of
filaments across field lines.

In section~\ref{sec:model} we model a loop observed by TRACE as a
bundle of thin parallel hydrostatic filaments, synthesize its emission in
the relevant TRACE bands and derive its effective temperature and
emission measure with the procedures used to analyze real TRACE data.
We discuss the results in section~\ref{sec:discuss}.

\section{The modeling}
\label{sec:model}

We consider a loop of half length $5 \times 10^9$ cm (corresponding to
loop c in L99). The choice of the shortest loop allows us to
concentrate on the effects of loop filamentation, without the
complication of a strong gravitational stratification expected for the
longer loops.

We model the loop as a bundle of parallel, static and hydrostatic
filaments.  Each filament is assumed to be semicircular and symmetric
with respect to the apex, and to lie on a plane perpendicular to the
solar surface.  Temperature and density stratification is taken into
account, according to Serio et al. (1981), by solving numerically the
equations for hydrostatic equilibrium and for energy balance among
plasma thermal conduction, radiative losses and a heating source
assumed uniform along each filament\footnote{Serio et al. 1981 have
shown that non-uniform heating distributions do not change dramatically
the model results, and, in particular, the temperature stratification,
due to the effectiveness of thermal conduction, unless the heating is
very localized.}.

Table~\ref{tab:models} shows the relevant parameters of the model loop
filaments, all of the same length and differing for the heating rate.
The latter determines the plasma pressure conditions inside each
filament (according to the scaling laws of Rosner et al.  1978 and of
Serio et al.  1981).  We have considered six filaments, with heating
rates such that the pressures at the footpoints are logarithmically
equispaced and spanning between 0.03 and 10 dyn cm$^{-2}$. The
corresponding filament maximum temperatures span between 0.8 and 5.2
MK. With this choice we sample the plasma conditions typical of non-flaring
coronal loops, from quiet to active regions, and relevant for observations
with TRACE 171 \AA~ and 195 \AA~ filters.

\placetable{tab:models}

\begin{figure}
\figurenum{1}
\plotone{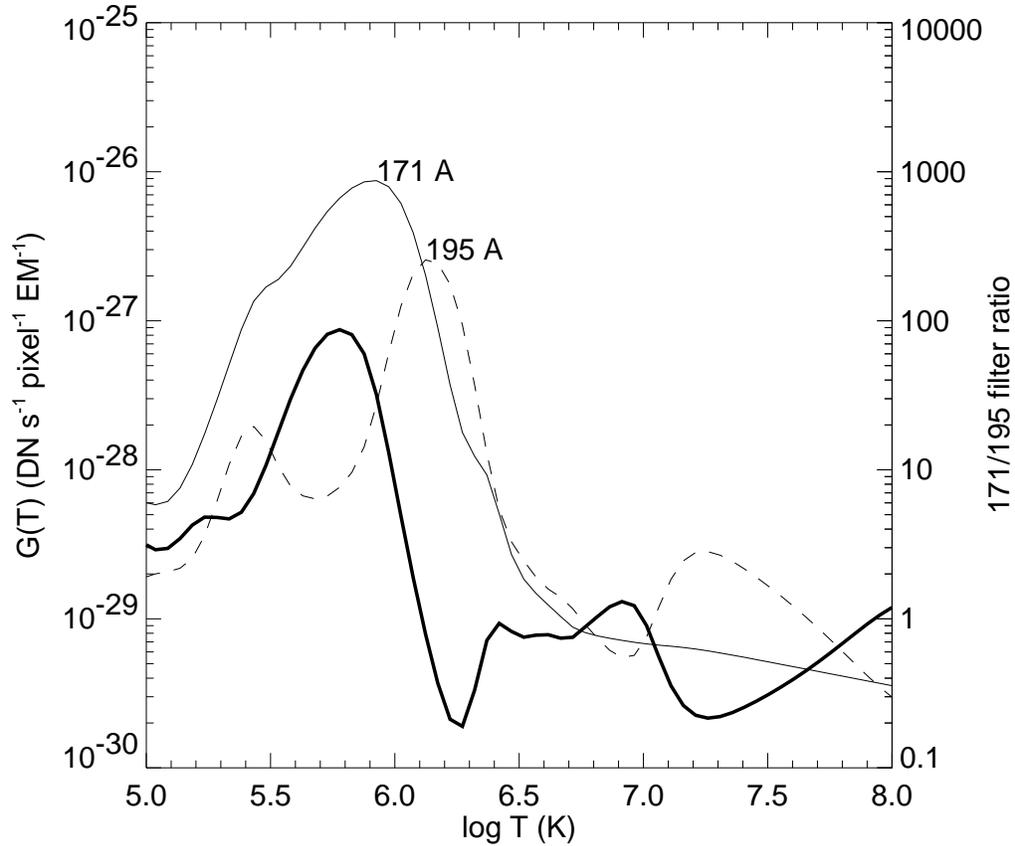}
\figcaption[fig1]{TRACE response of 171 \AA~ (solid) and 195 \AA~
(dashed) passbands (EM is in units of cm$^{-5}$), and filter ratio
(thick solid) of the 171 \AA~ and 195 \AA~ passbands vs temperature of
the emitting plasma volume.  \label{fig:gt+r}}
\end{figure}

The computed density and temperature profiles along half of each
filament (the other half is symmetric) have been used to synthesize 
the emission ${\cal E}$ per unit optical depth along the filament,
in units of Data Number (DN) s$^{-1}$ pix$^{-1}$ cm$^{-1}$,
in the two relevant TRACE passbands, according to:

\begin{equation}
{\cal E}(s) = n(s)^2 ~ G[T(s)]
\label{eq:emiss} 
\end{equation}
where $s$ is the coordinate along the filament (cm), $n$ the plasma
density (cm$^{-3}$), $T$ its temperature (K), $G(T)$ the response in
each TRACE passband (DN s$^{-1}$ pix$^{-1}$ EM$^{-1}$), computed with
the MEKAL spectral code (Mewe et al. 1995), and shown in
Fig.~\ref{fig:gt+r}. The figure also shows the expected ratio of the
emission detected in the 171 \AA~ and 195 \AA~ passbands vs temperature
(assuming ionization equilibrium).

\placefigure{fig:gt+r}

\begin{figure}
\figurenum{2}
\plotone{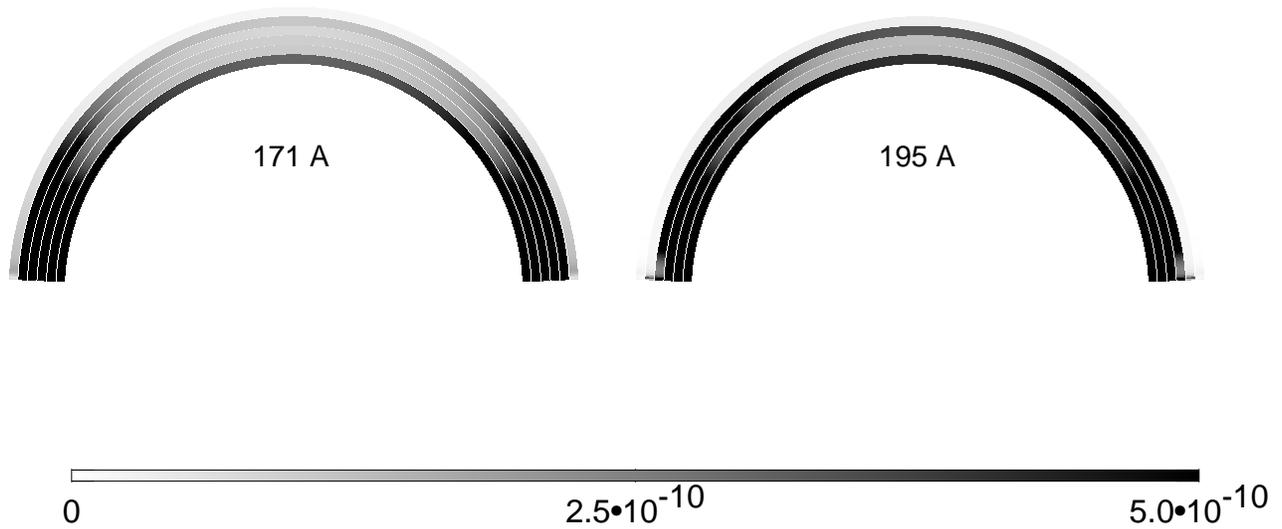}
\figcaption[fig2]{The loop-shaped structures show the distribution of
the emission per unit volume in the TRACE 171 \AA~ and 195 \AA~
passbands, synthesized from the six model loops in
Table~\ref{tab:models}, ordered with pressure (and maximum temperature)
increasing inwards. The grey scale (emission per unit optical depth) is
saturated at $5 \times 10^{-10}$ DN s$^{-1}$ pix$^{-1}$ cm$^{-1}$.
\label{fig:tvloops}}
\end{figure}

The loop-shaped pictures in Fig.~\ref{fig:tvloops} show the predicted
distributions of the emission in the two TRACE passbands for the six
filaments in Table~\ref{tab:models}, put side-by-side and ordered with
pressure (and maximum temperature) increasing inwards.  The grey scale
(emission per unit optical depth) is saturated at $5 \times 10^{-10}$
DN s$^{-1}$ pix$^{-1}$ cm$^{-1}$.

\placefigure{fig:tvloops}

Such pictures, which assume all filaments to have equal cross-section,
provide a general impression of how a thermally structured bundle of
filaments would look like in the TRACE passbands.  The outermost (and
coldest) filament is barely visible in the 171 \AA~ passband, even if
most of it is at a temperature not far from the peak of the response,
because of its low emission measure, and analogously invisible in the
195 \AA~ passband.  In the 171 \AA~ passband the hotter filaments have
comparable brightness in the chosen scale, except the hottest one which
is clearly brighter, because of its higher emission measure. In the 195
\AA~ passband filament 3 is brighter than filaments 2 and 4, because
the temperature of most of its plasma is close to the peak of the
filter response. Again, the hottest filament is the brightest one
because of its high emission measure.

\begin{figure}
\figurenum{3}
\plotone{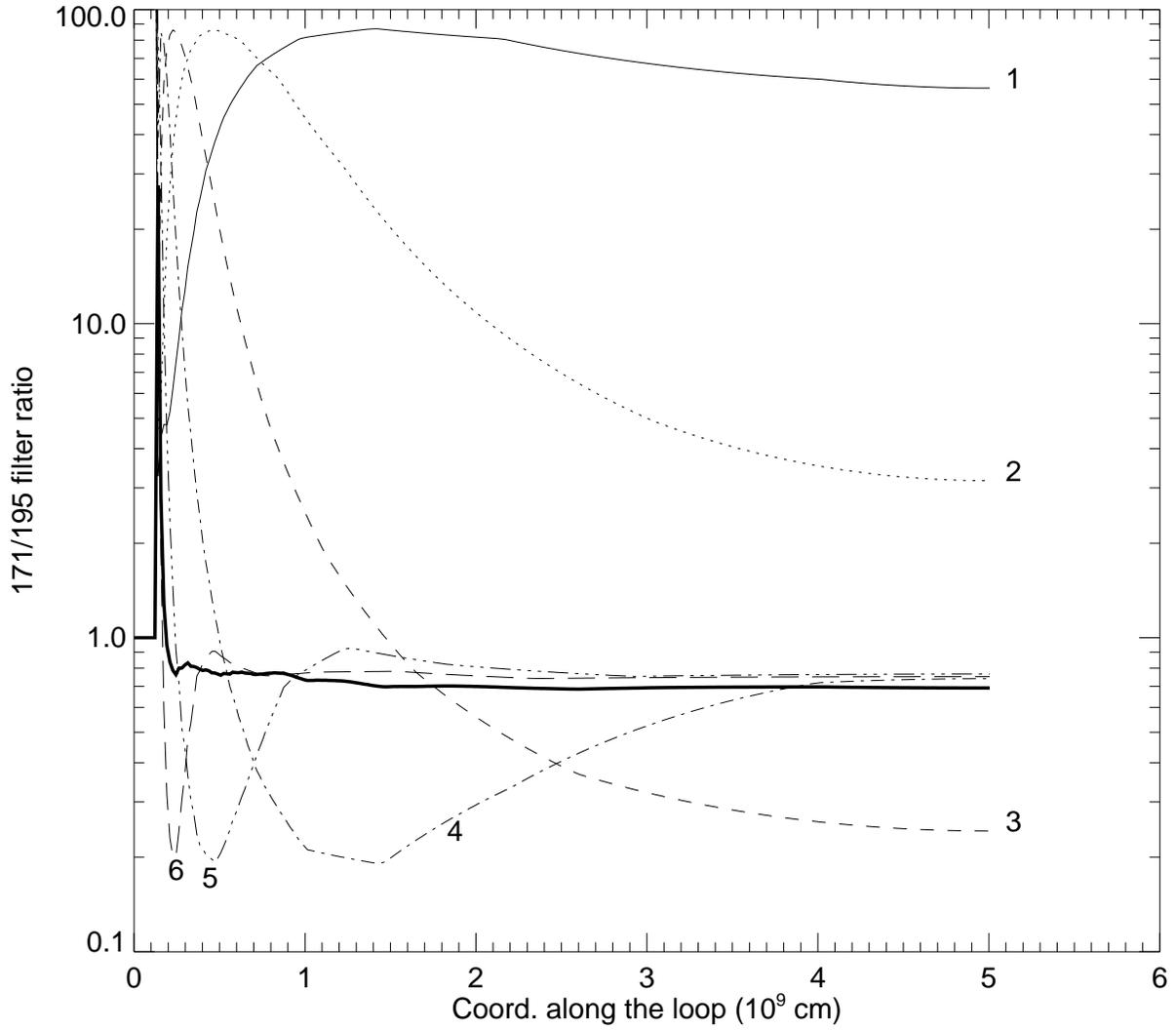}
\figcaption[fig3]{Ratio of the of the 171 \AA~ to the 195 \AA~ passband
expected emission computed along each of the six model loops in
Table~\ref{tab:models} (thin lines). The ratio obtained by merging the
six loops into a single loop, i.e. by summing the emission of the loops
is also shown (thick solid line).  \label{fig:ratio}}
\end{figure}

\placefigure{fig:ratio}

Fig.~\ref{fig:ratio} shows the ratio of the filtered emission in the
171 \AA~ to the 195 \AA~ passband computed along (half of) each of the
six filaments, from the footpoint to the apex, according to
Eq.~(\ref{eq:emiss}).  These profiles can be easily understood in terms
of the filter ratio curve shown in Fig.~\ref{fig:gt+r}.  Above the
lowest $10^9$ cm, the ratio is quite high for the coolest models 1 and
2 ($\gg 10$ for model 1), because the temperature along most of them is
in the branch below $\log T = 6.2$, and it is much lower than one along
most of filament 3 whose temperature is mostly around the ratio minimum
occurring at $\log T \approx 6.3$. The temperature of the other hotter
models mostly falls in the branch of the ratio curve above $\log T
\approx 6.3$, which is more weakly dependent on temperature and with a
value $\approx 1$.  The ratio profile of model 4 has a minimum close to
$10^9$~cm (corresponding to the minimum at $\log T \approx 6.3$ in the
ratio curve of Fig.~\ref{fig:gt+r}), and then gradually increases to
$\sim 1$ upwards along the loop. For models 5 and 6, the hottest ones,
the minimum is more localized and located well below the level of
$10^9$ cm: above, the ratio is constantly $\sim 1$.

Our key point is now how such a bundle of filaments would be detected
by TRACE if analyzed as a single loop, as in L99.  We then sum all the
emission profiles shown in Fig.~\ref{fig:tvloops}, to obtain two
single profiles, one for each filter passband.  The profile of the ratio of
the two resulting emission distributions is shown in
Fig.~\ref{fig:ratio} (thick solid line): the ratio is virtually
constant ($\approx 0.7$) along the whole loop except at the very
base (below $5 \times 10^8$ cm).  Of course, such a ratio is determined
mostly by the brightest filaments, and in particular model 6, the
hottest one, which contributes 30\% of the total emission above $10^9$
cm in the 171 \AA~ band; the faintest models 1 to 5 contribute to lower
the value of the ratio and to make it even more uniform.

\begin{figure}
\figurenum{4}
\plotone{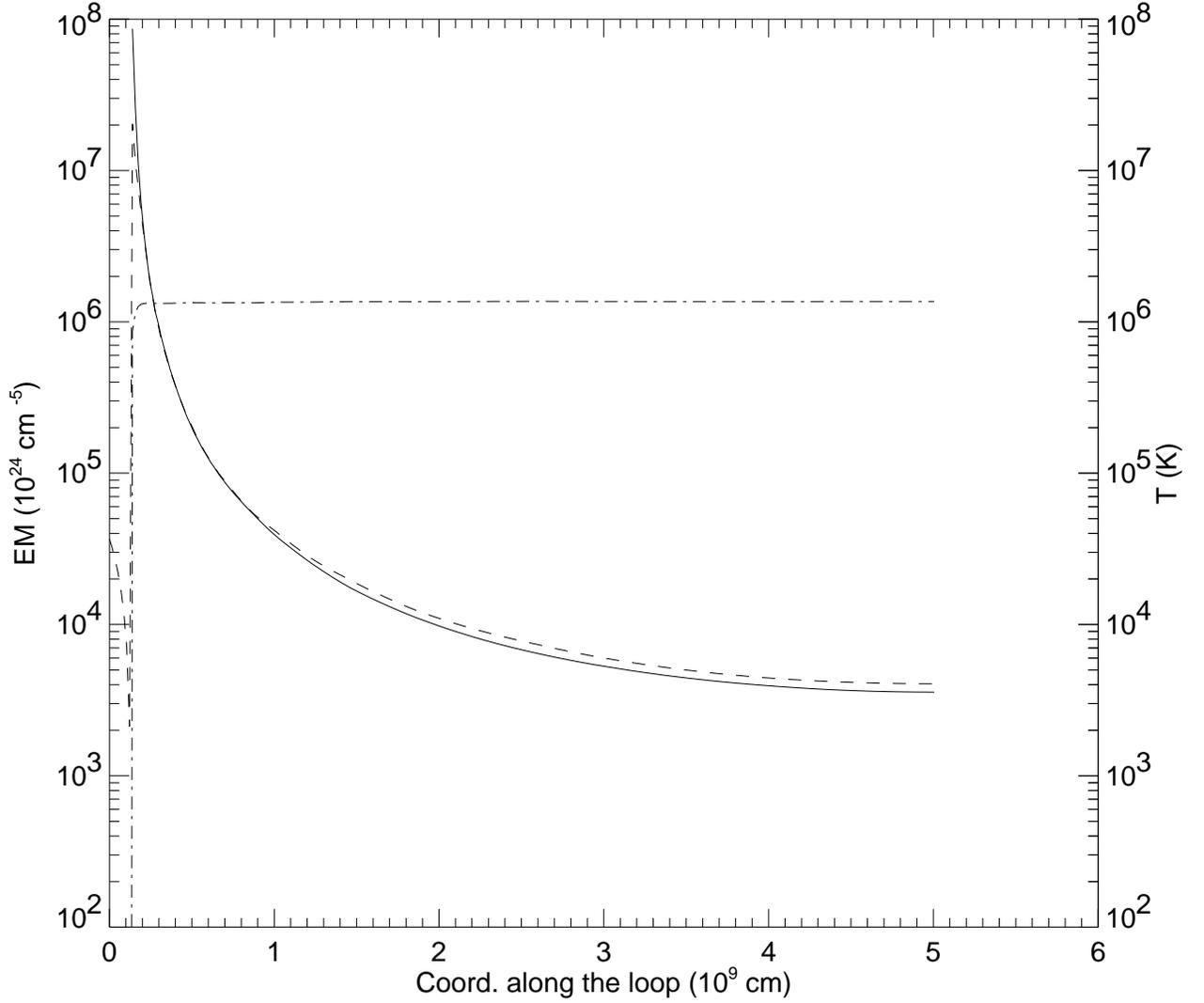}
\figcaption[fig4]{Temperature (dashed-dotted line) and emission measure
profiles along the bundle of six filaments (Table~\ref{tab:models}); we
have summed the synthesized emission of all six filaments in the 171
\AA~ and 195 \AA~ passbands and used the monotonic branch of the filter
ratio curve around $\log T  = 6.1$. The solid line shows the emission
measure profile derived, as in L99, using a constant G(T) value of $2
\times 10^{-27}$ DN s$^{-1}$ pix$^{-1}$ EM$^{-1}$ for the
171~\AA~passband; the dashed line shows the analogous result using the
G(T) proper.  \label{fig:em_t}}

\end{figure}

The most straightforward interpretation of such a flat ratio profile
would be a temperature uniformly distributed along the loop, and in the
region around $\log T = 6.1$, i.e. close to the temperature of maximum
formation of the 171 \AA~ and 195 \AA~ lines, if one restricts the
range of possible temperatures to be in the monotonic branch of the
filter ratio curve around $\log T  = 6.1$. The temperature
distribution, shown in Fig.~\ref{fig:em_t}, is flat for $s \geq 5
\times 10^8$ cm, and around 1.4~MK, i.e. very close to value obtained
by L99 for the four loops they have analized.

From the ratio of the expected emission to the value of the response
function at the given temperature, by assuming a line-of-sight depth of
$10^{10}$ cm, as done in L99, one obtains the distribution of the
emission measure shown in Fig.~\ref{fig:em_t}. The region of interest,
which can be compared to the results of L99, is the one above $10^9$
cm: in such 4/5 of the loop the emission measure spans between $\approx
4 \times 10^{28}$ cm$^{-5}$ and $\approx 4 \times 10^{27}$ cm$^{-5}$.
For loop c L99 report quite a flat profile, with a value around $5
\times 10^{27}$ cm$^{-5}$. The EM profiles shown in Fig.~\ref{fig:em_t}
are not flat above $10^9$ cm, but such profiles are obtained in the
assumption of a constant line-of-sight depth. This may not be entirely
realistic, as mentioned also in L99, and a smaller depth due to a
thinning of the loop near and at the chromosphere cannot be
excluded (a factor of two may be easily in agreement with
observations). If the shrinking is not very fast, this should not
affect significantly the model results, obtained in the assumption of a
constant cross-section. Note that the choice of the depth cannot
influence the key effect on the filter ratio discussed in this letter,
since it depends on the combination of the several filaments emission,
all six equally influenced by the depth value.

\placefigure{fig:em_t}

\section{Discussion}
\label{sec:discuss}

Base on the indication of isothermal loops, observed by TRACE, L99
conjecture the possibility of a new class of isothermal loops, distinct
from the non-isothermal hydrostatic loops.

The problem of the interpretation of the evidence of large isothermal
loops, although at temperatures lower than those in L99, and probably
in physical conditions very different from those in L99 (Brekke et al.
1997), have already been faced by Peres \& Orlando (1996) and Peres
(1997), who propose that such loops may be non-steady and strongly dynamic.

On the other hand, in the course of modeling a loop ignition observed
by TRACE, Reale et al. (1999) have found that the temperature profile
of a model loop of total length $10^{10}$ cm and heated at one
footpoint becomes relatively flat after a few thousand seconds.  Since
evidence for isothermality is invariably found in all loops analyzed in
L99, one wonders whether the phenomena discussed by Reale et al. (1999)
may be so common. Reale et al. (1999) themselves have found that the
loop they study is certainly not activated by heat deposition at one
footpoint, but more likely higher in corona. In the same work it is
shown that relatively flat temperature profiles are obtained if the
heating is high for a short time and then decaying slowly.  However,
such a possibility, as well as the non-steadiness proposed for large
cool loops, seems unlikely for the TRACE loops analyzed by L99, since
they are observed to be steady for times longer than the loop
characteristic cooling times.

The modeling illustrated in the present work shows that a bundle of
conventional (Serio et al. 1981) non-isothermal and uniformly-heated
static filaments with apex temperatures ranging between 0.8 and 5 MK,
observed by TRACE in the 171~\AA~ and 195~\AA~ passbands, if analyzed
as a single loop, would appear as an isothermal loop with a filter
ratio typical of $\sim 1.4$ MK and with emission measures in rough
agreement with the observed ones.

We have obtained this result by selecting a bundle of six loop
filaments with base pressure logarithmically equispaced, simply summing
their emission with equal weights.  One may wonder how the results are
influenced by this particular (although unbiased) choice of the
parameters; indeed we obtain very similar results with a different
choice of the pressures, provided that hot filaments are included. The
fundamental result is therefore quite robust.  Observations in all
three TRACE filters may provide further constraints on the thermal
structure of the loop bundle. However a more conclusive word may
require future instruments with even higher spatial resolution, so as to
better resolve the single filaments and to obtain enough
signal-to-noise ratio from each of them.

The key result obtained here crucially depends on the major role played
by the hottest components of the bundle: a) their temperature is mostly
in the region in which the 171/195 filter ratio is weakly dependent on
temperature, thus yielding a flat filter ratio profile along more than
4/5 of the length and showing a dip (corresponding to the dip in the
filter ratio curve around $\log T=6.3$) only very close to the
footpoints; b) their emission measure is high and contributes
significantly to the total emission in the TRACE band even though their
temperature is not close to the peak of the filter response.  For such
hot filaments, and, in general, for all loops with $6.3 < \log T < 7$,
the temperature diagnostics offered by TRACE two-filter observations
should be taken with care, due to the non-monotonic and weak dependence
of the 171/195 filter ratio on the temperature in that range.

Our conclusion is that the isothermal profiles obtained by L99 could be
a consequence of the high structuring of the observed loops across the
magnetic field lines, i.e.  ``filamentation", and, likely, an
indication of the presence of high temperature threads ($\sim 5$~MK).
The TRACE images, indeed also those in L99, show a high level of loop
filamentation.

This result of course needs further support by future observation and
analysis, possibly in coordination with other instruments with spectral
capabilities, such as those on board SOHO.

\bigskip
\bigskip
\acknowledgements{We thank S. Serio for useful suggestions.  We
acknowledge support from Ministero della Ricerca Scientifica e
Tecnologica and from Agenzia Spaziale Italiana.}

\onecolumn
\begin{table}
\begin{center}
\caption[]{Parameters of the model loop filaments\label{tab:models}}
\begin{tabular}{lcccc}
\hline
Model&Base pressure&$T_{max}$&Heating&Apex density\\
&(dyn cm$^{-2})$ &(MK)&($10^{-3}$ erg cm$^{-3}$ s$^{-1}$)& 
($10^{9}$ cm$^{-3}$)\\
\hline
1&0.03&0.8&0.01&0.05\\
2&0.1&1.1&0.04&0.15\\
3&0.3&1.6&0.15&0.4\\
4&1&2.4&0.6&1.1\\
5&3&3.5&2.6&2.5\\
6&10&5.2&10&5.9\\
\hline
\end{tabular}
\end{center}
\end{table}

\end{document}